\documentstyle[12pt,fleqn,epsfig]{article}
\def\beq {\begin{eqnarray}}
\def\eeq {\end{eqnarray}}
\def\beqn {\begin{eqnarray*}}
\def\eeqn {\end{eqnarray*}}
\def\neqn {\nonumber}

\def\ni {\noindent}
\def\new {\newpage}
\def\PL #1 #2 #3 {Phys. Lett.~{\bf#1} (#2) #3}
\def\NP #1 #2 #3 {Nucl. Phys.~{\bf#1} (#2) #3}
\def\ZP #1 #2 #3 {Z.~Phys.~{\bf#1} (#2) #3}
\def\PR #1 #2 #3 {Phys. Rev.~{\bf#1} (#2) #3}
\def\PP #1 #2 #3 {Phys. Rep.~{\bf#1} (#2) #3}
\def\PRL #1 #2 #3 {Phys. Rev.~Lett.~{\bf#1} (#2) #3}
\def\PTP #1 #2 #3 {Prog. Theor.~Phys.~{\bf#1} (#2) #3}
\def\MPL #1 #2 #3 {Mod. Phys.~Lett.~{\bf#1} (#2) #3}
\def\IJM #1 #2 #3 {Int. J.~Mod.~Phys.~{\bf#1} (#2) #3}
\def\etal {{\it et al}.}

\def\GeV{\mbox{GeV}}

\begin{document}
%\begin{titlepage}
%
\vspace* {-1cm}
\baselineskip= 1cm
\begin{center}{\large Analyzing powers
in inclusive pion production at high energy and \\the nucleon spin structure}

\baselineskip= 0.75cm
\vspace {1cm}
{\large
{K. Suzuki$^1$\footnote{e-mail address : ksuzuki@rcnp.osaka-u.ac.jp}, 
N. Nakajima$^1$, 
H. Toki$^1$ and
K.-I. Kubo$^2$}
}

$^1${\em Research Center for Nuclear Physics, Osaka University}\\
{\em Osaka 567-0047, Japan}\\
\vskip 0.4cm

$^2${\em Department of Physics, Tokyo Metropolitan University}\\
{\em Tokyo 192-0397, Japan}

\vspace {1cm}

{\bf Abstract}

\end{center}

\baselineskip= 0.7cm

Analyzing powers in inclusive pion production in 
high energy transversely polarized proton-proton collisions are studied
theoretically in the framework of the quark recombination model.  
Calculations by assuming 
the SU(6) spin-flavor symmetry for the
nucleon structure disagree with the experiments.  
We solve this difficulty by taking into account the
%We overcome this difficulty by taking into account the
realistic spin distribution functions of the 
nucleon, which differs from the SU(6) expectation at large $x$, 
%but coincides 
with a perturbative QCD constraint on the ratio of the 
unpolarized valence
distributions, $u/d \rightarrow 5$ as $x \rightarrow 1$. 
We also discuss the kaon spin asymmetry and find $A_N(K^+) =  -A_N(K^0)$ 
in the polarized proton-proton collisions at large $x_F$.   

\vspace {1.2cm} 

\ni

\vspace{0.3cm}
\ni

%

%\end{titlepage}
%%%%%%%%%%%%%%%%%%%%%%%%%%%%%%%%%%%%%%%%%%%%%%%%%%%%%%%%%%%%%%%%%%%%%%
\baselineskip= 0.7cm
\new

Against a naive expectation that spin effects become less
important at high energy, the significant polarizations
in inclusive hyperon productions\cite{Exp_hyp} and the large analyzing 
powers in
a pion production from a transversely polarized nucleon\cite{Exp_pi}
have been observed at low transverse momentum
$p_T$ and high Feynman $x_F$$(= 2 p_L/ \sqrt{s}$ in CM, $p_L$ is the
longitudinal momentum of the observed hadron). 
Such unexpected spin phenomena have attracted considerable 
experimental and theoretical
interests\cite{DeGrand,Yamamoto,Theo_hyp,Theo_pi}.  
In ref.~\cite{Yamamoto}, Yamamoto et al.~constructed a simple 
relativistic model for recombinations of quarks and/or 
diquarks to produce a final state hadron in terms of
quark distribution functions of 
incident hadrons and wave functions of the final state hadron. 
In this model, polarizations and analyzing powers are generated by
the scalar confining color force through the hadronization process in
purely non-perturbative way.  
It was demonstrated
that this model reproduces the empirical rule of DeGrand and
Miettinen (DM) \cite{DeGrand}, and provides polarizations and
asymmetries in good 
agreement with experiments\cite{Yamamoto}.

In this brief report, we concentrate on the transverse single spin 
asymmetry (analyzing power) 
in $\vec p + p \rightarrow \pi^a + X$ at high $x_F$, where
$a$ denotes the three pion charge states, $a =  \pm , 0$.
The original DM rule as well as the result of ref.~\cite{Yamamoto}
predict the relative magnitudes of analyzing powers in $\pi^+$, 
$\pi^-$ and
$\pi^0$ as $A_N (\pi^+) : A_N (\pi^-) : A_N (\pi^0) =  2:-1:1$,
while the experimental data indicate
$A_N (\pi^+) : A_N (\pi^-) : A_N (\pi^0) \sim 2:-2:1$. 
It will be shown that this defect comes from the assumption of SU(6)
spin-flavor symmetry
for the quark spin structure of the nucleon.

Here, instead of using the SU(6) symmetry assumption, we shall take more 
realistic approach by considering the spin-dependent structure function
measured in the lepton-hadron deep inelastic scattering. 
In this recombination process,
the valence quark distributions of the proton at large Bjorken $x$,
which are nothing but probabilities to find fast moving quarks in the
proton, are essential quantities to determine the analyzing power,
since we are only interested in fast pions (high $x_F$) in
the forward direction. 
On the other hand, experimental data of the deep inelastic
scattering tell us that the large $x$ behavior of the quark distribution 
function shows a significant deviation from the SU(6) predictions. 
These observations naturally lead us to apply the realistic and 
reasonable spin distribution 
function of the nucleon to the study of the analyzing powers.

To be more precise, we outline the quark recombination model which
is designed to describe 
the inclusive particle production for low $p_T$ and high 
$x_F$\cite{QRC}. 
In this model, a fast valence quark from the incoming proton picks up
one of the slow antiquarks created by the collision
in order to form a final state pion. 
This difference of momenta of quark and antiquark is 
indispensable to induce the spin dependence of the production cross 
section.  The asymmetry would vanish if momenta of both quarks were  
equal.  
We adopt the following basic assumptions to generate the
non-vanishing analyzing powers;   \\
(1) The final state hadron is produced by the simple quark 
recombination
process, since the observed single spin asymmetry is
significant only at large $x_F$.  \\
(2) Each parton which
participates in this reaction has the intrinsic transverse momentum
distribution.\\
(3) Quarks and antiquarks are combined by the scalar
confinement interaction in the hadronization process.  \\
Details are found in ref.~\cite{Yamamoto}.  
Our model naturally accounts for the phenomenological rule 
developed by DeGrand and Miettinen\cite{DeGrand}, which reproduces 
the relative ratios of the existing hyperon polarization data very well.

By choosing the $x$-axis as the beam direction and $z$ as
the transverse spin orientation, the production probability of a 
pion from the proton 
in the infinite momentum frame (IMF) is given
by\cite{Yamamoto}
\beq
&&S_{\pi \, p}  =  \int [dx_i \; dy_i\; dz_i / x_i]
G_\pi (x_3,x_4,y_3,y_4,z_3,z_4) \neqn \\
&& \hspace{2.3cm} \times | M (x_i,y_i,z_i) |^2
G_2^{sea} (x_2,y_2,z_2) \; G_1^p (x_1,y_1,z_1) \;
\Delta ^4 \; \Delta^3
\eeq
where $G_1^p(x_1,y_2,z_3)$ is the quark wave functions of the incoming 
proton 
with $x_i$ being the longitudinal momentum fraction (0$ \le x_i \le 1$)
and $y_i, z_i$ the transverse momentum fractions.    
$G_2^{sea}$ denotes the momentum distribution of the slow picked-up 
antiquark developed by the non-perturbative hadronization process, and   
$G_\pi (x_3,x_4,y_3,y_4,z_3,z_4)$ is the 
light-cone pion wave function.  
$\Delta^3$ and $\Delta^4$ express the delta functions which correspond to the 
energy-momentum conservation in this process. 
$M$ represents the elementary hadronization amplitude $q_1 + \bar q_2 
\rightarrow q_3 + \bar q_4$ to produce the pion 
under the confining color field.  
We explicitly calculate these amplitudes using the scalar interaction.  
%The delta functions which respect energy-momentum conservation are 
%given by
%%
%\beq
%\Delta^3 =  \delta [x_F (x_4 + x_3-1)] \; \delta [y_4 + y_3 - P_T 
%/p_t] \;
%\delta [z_4 + z_3]
%\eeq
%
%\beq
%\Delta^4 &= & \delta [x_F (x_4 + x_3) - x_2 -x_1] \; \delta[y_4 + y_3 
%-y_2 -y_1] \;
%\delta [z_4 + z_3 - z_1 - z_2] \neqn \\
%&&\hspace{-2.5cm} \times \delta \left[\frac{(y_4^2+z_4^2)p_t^2 +
%m_2^2}{x_F x_4} +
%\frac{(y_3^2+z_3^2)p_t^2 + m_1^2}{x_F x_3} -
%\frac{(y_2^2+z_2^2)p_t^2 + m_2^2}{x_2} -
%\frac{(y_1^2+z_1^2)p_t^2 + m_1^2}{ x_1} \right]
%\label{delta}
%\eeq

For the quark wave function $G_1$ and $G_2$, we assume the following
factorized form;
\beq
G_i (x_i, y_i,z_i) =  q_i(x_i) \; e^{-y_i^2} \; e^{-z_i^2}
\label{p_dist}
\eeq
where $q_i(x)$ is the quark distribution function measured in the
deep inelastic
scattering, while we use the Gaussian momentum distribution for the 
transverse $y$ and $z$ components. 
Average value of the intrinsic
transverse momentum is assumed to be 300MeV\cite{Yamamoto}. 
On the other hand, we use the pion wave function based on the light-cone 
formalism, which is given by\cite{pion_light}
\beqn
G_\pi (x_3,x_4,k_t) = A \; \, \mbox{exp} \left[ - \frac{1}{8 \beta^2} 
\left[\frac{k_t^2 + m^2}{x_3} + \frac{k_t^2 + m^2}{x_4}\right] 
\right]   
\eeqn

Here, the transverse single spin asymmetry, analyzing power, 
is defined by, 
\beq
A_N (\pi) = \frac{d \sigma (\pi; \uparrow) - d \sigma (\pi; \downarrow)} 
{d \sigma (\pi; \uparrow) + d \sigma (\pi; \downarrow)} 
\eeq
where $d \sigma (\pi; \uparrow (\downarrow))$ means the pion production 
cross section with the spin direction of the beam proton being $+z (-z)$.  
We arrive at an expression to obtain the analyzing power;
\beq
A_N (\pi) &= &R \frac{  \int [dx_i \; dy_i\; dz_i / x_i]
G_\pi \; \sigma_{dep} \;
G_2^{sea} \; (G_1^{p \uparrow} - G_1^{p \downarrow}) \; \Delta^4 
\Delta^3   }
{\int [dx_i \; dy_i\; dz_i / x_i]
G_\pi \; \sigma_{ind} \;
G_2^{sea} \; G_1^p \; \Delta^4 \Delta^3}
%\neqn \\
%&= &R \frac{  \int [dx_i \; dy_i\; dz_i / x_i]
%G_4^\pi  \; G_3^\pi \; \sigma_{dep} \;
%G_2^{sea} \; \delta G_1^{p} \; \Delta^4 \Delta^3 }
%{\int [dx_i \; dy_i\; dz_i / x_i]
%G_4^\pi  \; G_3^\pi \; \sigma_{ind} \;
%$G_2^{sea} \; G_1^p \; \Delta^4 \Delta^3}  
\; \; ,
\label{ana}
\eeq
where the spin-independent cross section is 
\beqn
&&\sigma_{ind} =  (x_F x_4 x_2) (x_F x_3 x_1)
\left[ \left( \frac{x_F x_4 + x_2}{x_F x_4 x_2} m_2  \right)^2 +
\left(\frac{x_F x_4 y_2 - x_2 y_4}{x_F x_4 x_2} \bar p_t \right)^2 \right]
\\
&&\hspace{2cm} \times
\left[ \left(\frac{x_F x_3 + x_1}{x_F x_3 x_1} m_1 \right)^2 +
\left(\frac{x_F x_3 y_1 - x_1 y_3}{x_F x_3 x_1} \bar p_t \right)^2 
\right] \; ,
\eeqn
and the spin-dependent one 
\beqn
&&\sigma_{dep} =  (x_F x_4 x_2) (x_F x_3 x_1)
\left[ \left(\frac{x_F x_4 + x_2}{x_F x_4 x_2} m_2 \right)
\left(\frac{x_F x_3 y_1 - x_1 y_3}{x_F x_3 x_1} \bar p_t \right)
\right. \\
&& \hspace{2cm} - \left.
\left(\frac{x_F x_3 + x_1}{x_F x_3 x_1} m_1 \right)
\left(\frac{x_F x_4 y_2 - x_2 y_4}{x_F x_4 x_2} \bar p_t \right) 
\right] \; \; .
\eeqn
$R$ involves `unknown' underlying dynamics of the confinement force, and  
is simply assumed to be a constant parameter which will be fixed 
to reproduce the $\pi^+$ analyzing power.  
Note that the spin dependent part of the cross section $\sigma_{dep}$
appears as a result of the interference between the leading 
order diagram and higher order one in the 
non-perturbative hadronization process\cite{Yamamoto}.  
It is important to note that, if we took the vector type interaction instead 
of the scalar, the resulting spin dependent cross section would vanish 
in the IMF.

The spin dependent momentum distribution function,
$\delta G =  G^\uparrow -G^\downarrow$,
of the proton is defined by
\beq
\delta G_1 (x_1,y_1,z_1)  =  \delta q_1(x_1) e^{-y_1^2} e^{-z_1^2}
\label{nucl_dis}
\eeq
where $\delta q_1(x)$ is the spin dependent quark distribution function
as a function of the longitudinal momentum fraction $x_1$.

The $\pi^+$ production process is dominated by the valence up-quark in
the proton, 
$q_1 (x_1) =  u(x_1)$, and $\pi^-$ case by the valence down-quark,
$q_1 (x_1) =  d(x_1)$, in the present kinematical region. 
Since we take a ratio in eq.~(\ref{ana}),
resulting analyzing powers are rather insensitive
to the shapes of the wave functions $G_2^{sea}$ and  $G_\pi$. 
The momentum conservation requires $\delta (x_F(x_4 + x_3) -x_2-x_1)$ in 
eq.~(\ref{ana}).   
It is easy to understand that, 
in order to produce fast pions (high $x_F$), $x_1$ must be large,
because the {\em slow} antiquark distribution $G_2^{sea} (x_2) $
has a peak at relatively small momentum
fraction,  $x_2 \ll 1$\cite{QRC}.  
%\footnote{Note that the antiquark 
%distribution
%$G_2^{sea}$ cannot be simply identified with the sea quark distributions 
%of the
%target nucleon, because a lot of slow quarks and antiquarks have been 
%produced
%by the proton-proton collision before the recombination process.
%A great number of $q \bar q$ pairs are also created through the
%hadronization. $G_2^{sea}$ must involve all these contributions.}. 
Therefore, we point out that the large Bjorken $x$
behavior of the
quark distribution functions of the incoming proton, $q_1(x_1)$ and
$\delta q_1(x_1)$, mainly controles the analyzing power of pions. 
In other words, $A_N (\pi^+)$ is sensitive to the shape of
$\delta u(x) / u(x)$
of the proton, and $A_N(\pi^-)$ to $\delta d(x) / d(x)$.

If we assume the SU(6) spin-flavor symmetry for the nucleon,
the spin
distribution function $\delta q_1(x)$ appeared in eq.~(\ref{nucl_dis})
are written by
\beq
\delta u(x) & =  & \frac{2}{3} u (x)\\
\delta d(x) & =  & - \frac{1}{3} d(x)
\eeq
with $u(x)=  2d(x)$ for unpolarized distribution functions.  
Inserting them into eq.~(\ref{ana}), one easily finds
\beqn
A_N(\pi^+): A_N(\pi^-) =  2:-1
\eeqn
This disagrees with experiments as we have already discussed.

It is well known from deep inelastic experiments that valence quark
distribution functions of the nucleon at large Bjorken $x$ are clearly
different from
the SU(6) symmetry expectations.  A ratio of neutron to proton
structure functions $F_2^n(x) / F_2^p(x)$ is much smaller than the
SU(6) value 2/3 at $x \sim 1$\cite{Exp_F2pn}. 
Similarly, the ratio of the spin-dependent to spin-independent structure 
functions 
of the proton $g_1^p(x) / f_1^p(x)$ approaches to 1 at $x \rightarrow 
1$
against the SU(6) value 5/9\cite{Exp_g1p}. 
These facts suggest that the SU(6) spin-flavor symmetry is not a
realistic assumption on the spin quantities
any more at large Bjorken $x$, at which one of the valence quarks
carries most of the nucleon momentum,
though the SU(6) symmetry may works
well for $x$-integrated moments of the structure functions.

Here, we introduce the spin dependent distribution function $\delta 
q(x)$
in the following simplified form to mimic the shapes of the experimental 
data on $g_1^p(x) / f_1^p(x)$\cite{Exp_g1p};
\beq
\frac{\delta u(x)}{u(x)} &= & \sqrt{x} \neqn \\
\frac{\delta d(x)}{d (x)} &= & - \sqrt{x}
\label{spin}
\eeq
Choices of eq.~(\ref{spin}) reasonably reproduce
the $x$-dependence of $g_1^p(x) / f_1^p(x)$\footnote{
Strictly speaking, spin distribution function which we need here is the
`transversity' distribution function in a transversely polarized
nucleon\cite{JaffeJi}, and different from the `helicity'
distribution function measured in the
longitudinally polarized lepton-nucleon scattering. 
Nevertheless, in the first approximation, the transversity distribution 
and
the helicity distribution are considered to be very similar as expected 
by the naive
quark model.   Thus, we assume the behavior in eq.~(\ref{spin}).  
It can be shown that the choices of $\delta q(x)$ in eq.(\ref{spin})
does not violate the Soffer's inequality\cite{Soffer}
for the spin-independent, helicity and transversity
distribution functions.}. 
%%%%%%%%%%%%%%%%%%%%%%%%%%%%%%%%%%%%%%%%%%%%%%%%%%%%%% 
Quark spin fractions calculated by using eq.~(\ref{spin}) are found to 
be consistent with the present data. 
Such a behavior of the spin distribution function is also suggested by 
the several quark model calculations\cite{Suzuki,JaffeJi}.

In the previous discussions we have introduced
$\delta u(x)$ and $\delta d(x)$ that deviate from the SU(6) at $x 
\sim 1$. 
However, this prescription makes another subtle trouble for the
analyzing power in $\pi^0$ production. 
The asymmetry of $\pi^0$ in this model is given by,
\beq
A_N (\pi^0)
=  R \frac{  \int [dx_i \; dy_i\; dz_i / x_i]
\frac{ \delta u(x) + \delta d(x) }{2}  e^{-y_1^2} e^{-z_1^2} \; 
G_2^{sea}
\; \sigma_{dep} \; G_\pi  \; \Delta ^4 \Delta ^3}
{\int [dx_i \; dy_i\; dz_i / x_i]
\frac{  u(x) + d(x) }{2}  e^{-y_1^2} e^{-z_1^2} \; G_2^{sea}
\; \sigma_{ind} \; G_\pi \; \Delta ^4 \Delta ^3} \; \; . 
\label{annpi0}
\eeq
This expression indicates that the single spin asymmetry of $\pi^0$ is
governed by $ (\delta u(x) + \delta d(x) )/
(u(x) + d(x))$ at large $x$. 
In the standard parametrization of valence quark distribution
functions of the proton \cite{CTEQ,MRST} extracted from the available
lepton-hadron scattering, it is assumed that
\beq
d(x) / u(x) \rightarrow 0
\label{cteq-ratio}
\eeq
at large $x$.  Consequently, only the up quark distribution survives at
the large $x$,  and eq.~(\ref{annpi0}) at high $x_F$
can be rewritten as
\beq
A_N (\pi^0)
\sim  R \frac{  \int [dx_i \; dy_i\; dz_i / x_i]
\delta u(x)   e^{-y_1^2} e^{-z_1^2} \; G_2^{sea}
\; \sigma_{dep} \; G_\pi \; \Delta ^4 \Delta ^3 }
{\int [dx_i \; dy_i\; dz_i / x_i]
u(x)   e^{-y_1^2} e^{-z_1^2} \; G_2^{sea}
\; \sigma_{ind} \; G_\pi \; \Delta ^4 \Delta ^3 } \; . 
\eeq
This is the same as the $\pi^+$ case, and hence $A_N(\pi^+) =  A_N
(\pi^0)$, which disagrees with the experiments.

However, it was recently pointed out that there still exist ambiguities
to extract the large $x$ behavior of the unpolarized 
quark distribution functions in the deuteron, from which we determine the 
flavor dependence of the quark distributions.     
Melnitchouk and Thomas \cite{Mel_Thom} discussed that, using the recent
developments  to treat nuclear effects in the deuteron,
present experimental data
are shown to be rather consistent with the result of the perturbative
QCD\cite{ud}
\beq
d(x) / u(x) \rightarrow 1/5
\label{dist-ud}
\eeq
as $x \rightarrow 1$. 
Here, we shall adopt this constraint for $u$ and $d$ unpolarized
quark distributions. 
We will show that this behavior is crucial to account for
both the unpolarized cross section and analyzing powers.

In practice, we shall fix the input quark distribution functions of the
proton  as follows. 
For the unpolarized distribution function, we basically use the CTEQ4
parametrization of the quark distribution functions\cite{CTEQ}. 
We modify it to keep a constraint  $d(x) / u(x)$$\rightarrow 1/5$
at $x \rightarrow 1$, instead of the original CTEQ4 where $d(x) / u(x)
\rightarrow 0$. 
Our unpolarized quark distribution functions are shown in Fig.1 with
the original CTEQ4 distributions. 
We cannot see any sizable difference for the
$u$-quark, but slight increase of the $d$-quark distribution at
large-$x$.  Obtained shapes of these distributions are
consistent with
the recent analysis of ref.~\cite{Unpol}, in which the nuclear binding 
effects of the deuteron are taken into account.  
Using the unpolarized distributions and multiplying them by factors in
eq.~(\ref{spin}), we obtain the spin dependent
distribution functions $\delta q(x)$.

\begin{figure}[htb]
\begin{center}
\psfig{file=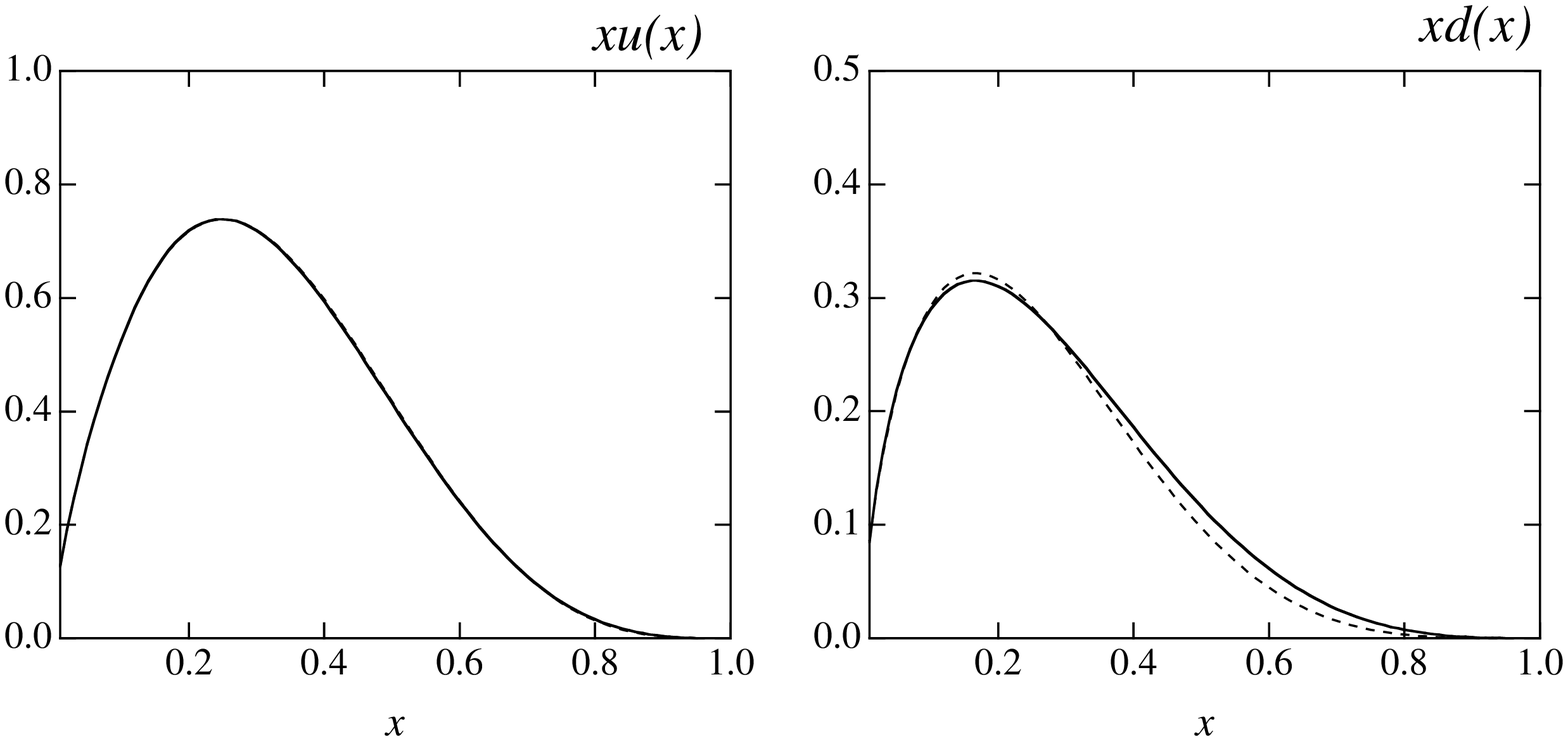,height=3.5in} 
\vspace{-2cm}
\caption{Unpolarized valence quark distribution functions of the proton at
$Q^2 =  1 \GeV^2$. 
The $u$-quark distribution is depicted in the left figure, and $d$-quark
in the right one. 
Our parametrizations are shown by the solid curves and
CTEQ4LQ\cite{CTEQ} by the dashed curves.}
\label{fig1}
\end{center}
\end{figure}

We also use the following distribution functions to get numerical results.  
The transverse momentum distributions for $y$ and $z$ components are 
assumed to be the Gaussian types as already introduced in 
eq.~(\ref{p_dist}).   
The picked-up antiquark distribution function is taken to be 
$g_2^{sea} (x) \sim C (1-x)^4 x^{-0.5}$ ($C$ is the constant number).  
Compared with the standard parametrization 
of the antiquark distribution of the proton\cite{CTEQ}, $g_2^{sea}(x)$ 
possibly involves high momentum components, because we assume that 
this antiquark distribution originates from the $q \bar q$ pair 
creation by the 
string breaking in the soft hadronization process.   
%and also the antiquarks 
%produced in the initial hard proton-proton collision.   
The resulting analyzing powers are insensitive to the variation of 
$g_2^{sea}$ and  $G_\pi (x,y,z)$.   
For example, even if we take $g_2^{sea} \sim (1-x)^{6-8}$ adopted in 
ref.~\cite{QRC}, the behavior of the analyzing power is
almost unchanged.   
It may be possible to fix the picked-up antiquark distribution 
in order to reproduce the $x_F$ and $p_T$ dependence of the 
unpolarized cross section, and such a study is in progress\cite{Nakajima}.

Before we shall discuss the analyzing powers, let us consider the
unpolarized cross section for $\pi^+$ and $\pi^-$ at large $x_F$ in the
quark recombination model.  Parameters of the model are already fixed to 
reproduce the data of the hyperon polarizations, which can be found in
ref.~\cite{Yamamoto}. 
In Fig.2 we show a ratio of $\pi^+$ to $\pi^-$ cross sections
with experiments\cite{Exp_ratio}. 
The calculation agrees with the data nicely.  If we used the CTEQ4
or other standard parton distribution functions where $u/d \rightarrow
\infty$ ($x \rightarrow 1$),  this curve would blow
up at $x_F \sim 1$, which seems to be inconsistent with the data.

\begin{figure}[htb]
\begin{center}
\psfig{file=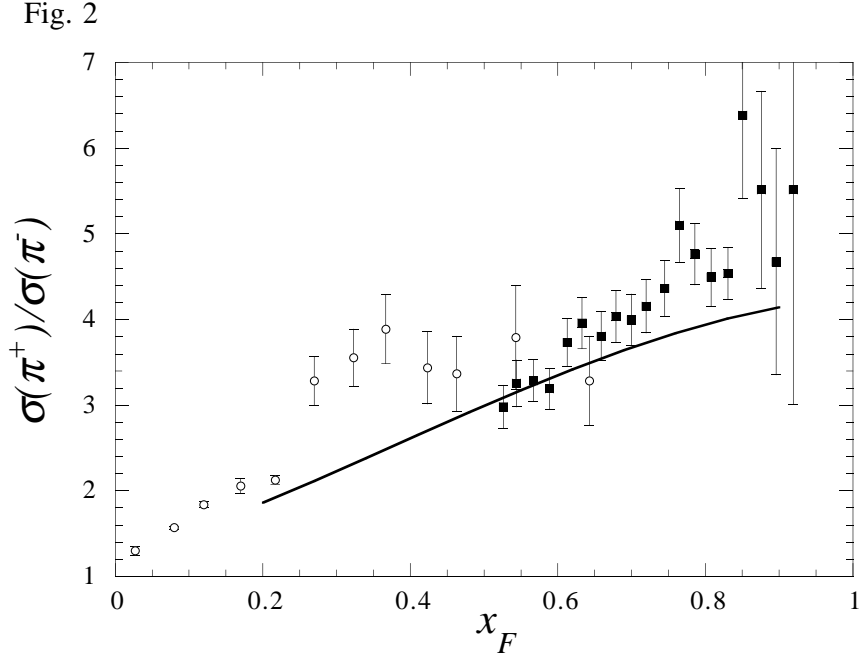,height=3.5in} 
\vspace{-0.4cm}
\caption{Ratio of the inclusive $\pi^+$ to $\pi^-$ production cross sections at
$p_T= 0.75 \GeV$.  Experimental data are taken from 
ref.~\cite{Exp_ratio}.}
\label{fig2}
\end{center}
\end{figure}

We finally show in Fig.3 spin asymmetries for pions with the
experimental data at $p_T = 0.75 \GeV$. 
Absolute magnitude of $A_N (\pi^-)$ is almost the same as one of
$A_N (\pi^+)$, which reasonably reproduces the data. 
The asymmetry of $\pi^0$ is also in a good agreement. 
These results are understood by the following very simplified 
discussion.
At $x_F \sim 1$, relative magnitudes of the analyzing powers in  $\pi^a$ 
are intuitively given by
ratios of spin-dependent to spin-independent distribution functions at
$x \sim 1$,
\beqn
A_N(\pi^+) : A_N(\pi^-) : A_N(\pi^0)
\sim \frac {\delta u(x \sim 1)} {u(x \sim 1)} :  \frac {\delta d(x
\sim 1)} {d(x \sim 1)} :
\frac {\delta u(x \sim 1) + \delta d(x \sim 1)}
{u(x \sim 1) + d(x \sim 1)} \; .
\eeqn
According to our procedure in eqs.~(\ref{spin},\ref{dist-ud}), we
employ a relation $u(x) =  5d(x)$
and $\delta u(x) =  u(x) $, $\delta d(x) =  -d(x)$ at $x \sim 1$. 
Thus, it is easy to see,
\beq
A_N(\pi^+):A_N(\pi^-):A_N(\pi^0) \sim  1:-1: \frac{2}{3}
\hspace{1cm} \mbox{at } \hspace{0.3cm} x_F \sim 1 \; ,
\eeq
which evidently accounts for essential features of the experiments shown 
in Fig.3. 
Also, our calculations reasonably describe the data on the pion analyzing
powers of the polarized antiproton-proton scatttering\cite{Exp_pi}.

\begin{figure}[htb]
\begin{center}
\psfig{file=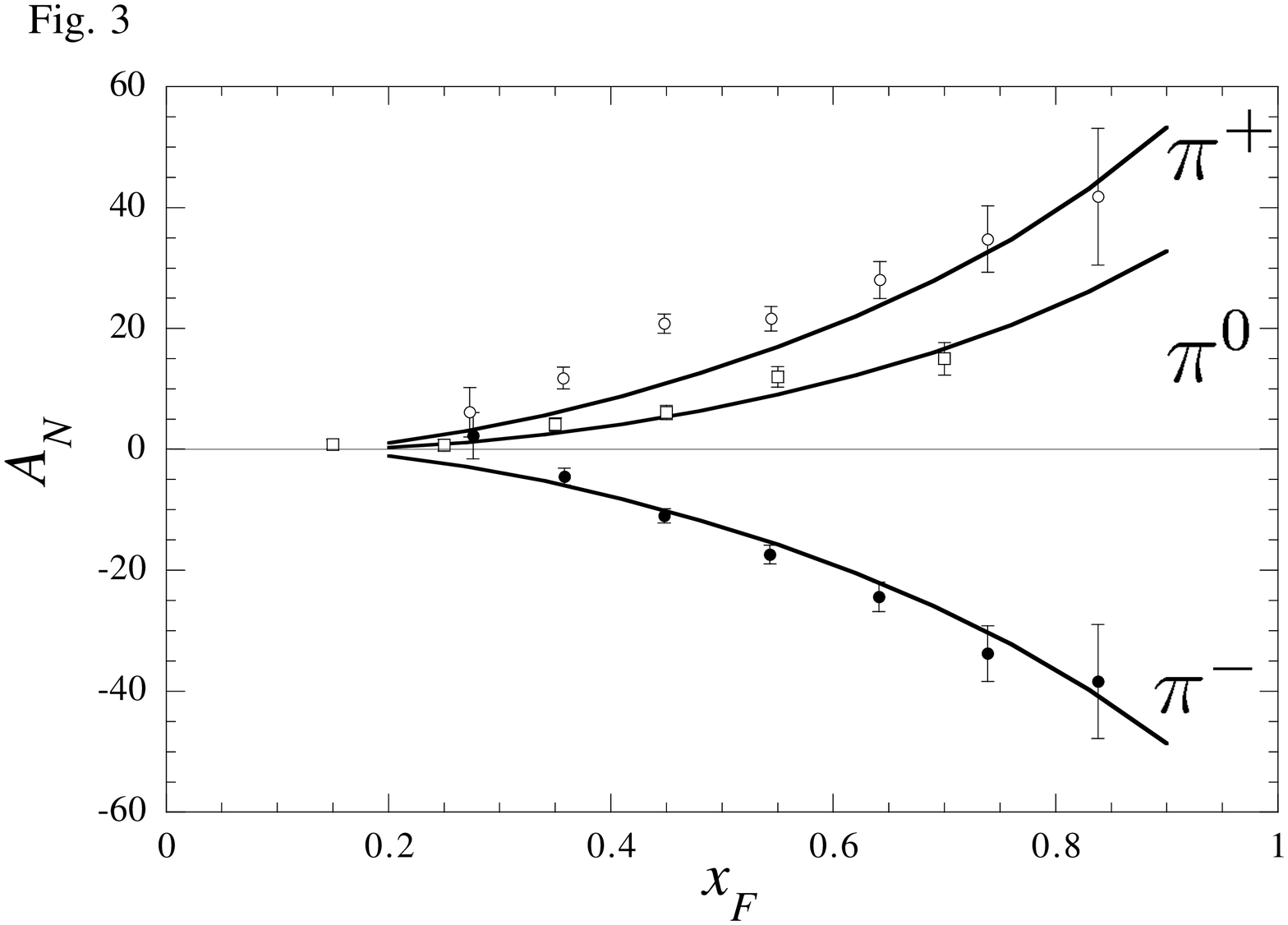,height=3.5in} 
\vspace{-0.4cm}
\caption{Analyzing powers of $\pi^+$, $\pi^-$ and $\pi^0$ from the transversely
polarized proton with $p_T =  0.75 \GeV$.  Experimental data are 
depicted by
the open circles, filled circles and open boxes for
$\pi^+$, $\pi^-$ and $\pi^0$, respectively\cite{Exp_pi}. 
Our results are shown by the solid curves.}
\label{fig3}
\end{center}
\end{figure}

To be more realistic, {\em true} behavior of $u(x) / d(x) $ lies somewhere 
between the standard 
prametrization eq.~(\ref{cteq-ratio}) 
($u(x) / d(x) \to \infty $) and the perturvative QCD 
motivated constraint eq.~(\ref{dist-ud}) ($u(x) / d(x) \to 5$).  
It is actually seen that 
the experimental data on the  
unpolarized cross section ratio $\sigma (\pi^+) / \sigma (\pi^-) 
\sim u(x) / d(x) $ is 
slightly larger than 5 obtained by the 
parametrization eq.~(\ref{dist-ud}).   
Anyways, the parametrization eq.~(\ref{dist-ud}) gives much 
better 
results for both the cross section ratio and the spin asymmetry.  
Therefore, we 
expect the perturbative QCD inspired distributions 
eq.~(\ref{dist-ud}) to be more reasonable as the 
valence quark distribution of the nucleon 
in this work.   
At the present, the experimental data of the analyzing power for $\pi^0$ 
are available at $x_F \leq 0.7$.  
Forthcoming data at much higher $x_F$ may provide some new constraints on the 
large-$x$ behavir of the quark distribution in the nucleon.

%As for $P_T$-dependence of the analyzing powers, which increase 
%linearly with
%$P_T$ increasing, our model basically explains the experimental result.
%In particular, the transverse momentum $P_T$ dependence of the
%analyzing powers

Analyzing powers in inclusive kaon production $\vec p + p
\rightarrow K + X$  can be calculated in the same
way.  This model predicts asymmetries in $ K^-$ and $\bar K^0$ vanish,
because the proton does not contains (fast)
valence $\bar u$ or $\bar d$ quarks. 
$K^+$ analyzing power is positive, whereas $K^0$ gives a negative 
asymmetry.  
We present the analying powers of $K^+$ and $K_0$ in Fig.4 with 
the kaon distribution\cite{Shigetani}.  
Relative magnitude for large $x_F$ kaons is obtained as
\beq
A_N(K^+) : A_N(K^0) =  1:-1 \; \; . 
\eeq
Note that the SU(6) symmetry consideration leads to a result
$A_N(K^+) : A_N(K^0) =  2:-1$.  This prediction will be tested
in future experiments. 

\begin{figure}[htb]
\begin{center}
\psfig{file=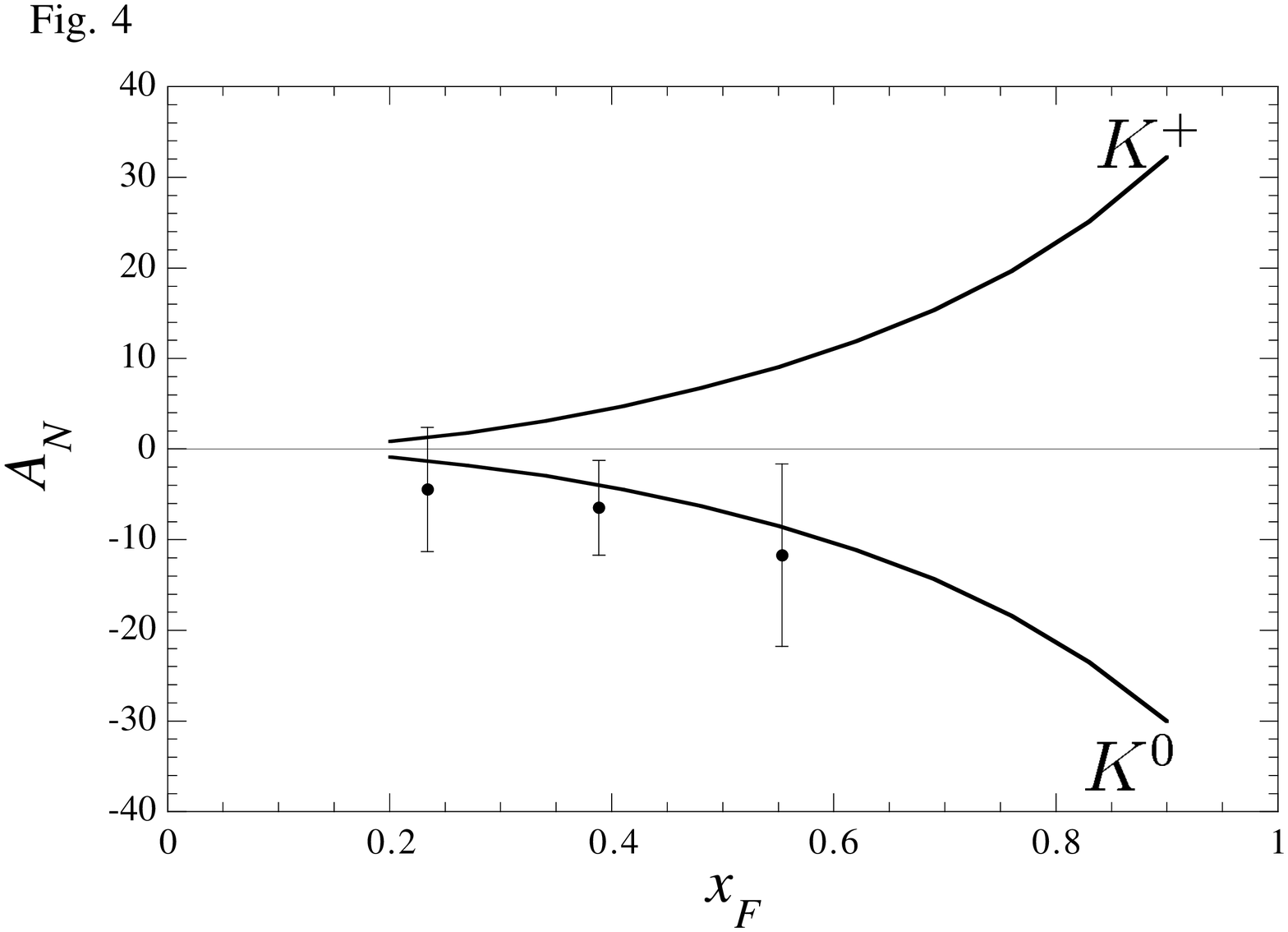,height=3.5in} 
\vspace{-0.4cm}
\caption{Analyzing powers of $K^+$ and $K^0$ with $p_T =  0.75 \GeV$
are shown by the solid and dashed curves, respectively.  
Experimental data of $K^0_s$ are also shown by the filled circles.  
}
\label{fig4}
\end{center}
\end{figure}

We also calculate the case of inclusive $\rho$ meson production for
polarizations and analyzing powers.  Results strongly depend on the
choice of the $\rho$ meson wave function, which will be published
elsewhere\cite{Nakajima}.

In conclusion, we have studied the analyzing powers in inclusive
pion productions $\vec p + p \rightarrow
\pi^a + X$ at high $x_F$ in terms of  the quark recombination model. 
We have particularly emphasized that the analyzing power
is sensitive to the large Bjorken $x$ dependence of the spin 
distribution function in the nucleon. 
Calculations based on the SU(6) spin-flavor symmetry of the nucleon
cannot describe relative magnitudes of analyzing powers
in $\pi^+$, $\pi^-$ and $\pi^0$. 
However, once we take into account the realistic quark
distribution functions which deviate from the SU(6) predictions at
large Bjorken $x$,
as suggested by the deep inelastic scattering and effective quark model
calculations, resulting analyzing powers show reasonable agreement with 
the data. 
These results may indicate that the spin dependent inclusive hadron 
production
at the high $x_F$ region,
which are accessible at RHIC and HERA-N, is a complemental tool to probe 
the valence quark spin structure of the nucleon at large $x$.

\vspace{1cm}

\ni
{\bf Acknowledgments}

K.S. would like to thank the COE program, which enables him
to work out this project at RCNP.

%%%%%%%%%%%%%%%%%%%%%%%%% references %%%%%%%%%%%%%%%%%%%%%%%%%%%%%%%%%
%
%\newpage

%
%%%%%%%%%%%%%%%%%%%%%%%%%%%%%%%%%%%%%%%%%%%%%%%%%%%%%%%%%%%%%%%%%%
\end{document}